# A New Distance for High Level RNA Secondary Structure Comparison

Julien Allali and Marie-France Sagot

**Abstract**—We describe an algorithm for comparing two RNA secondary structures coded in the form of trees that introduces two new operations, called *node fusion* and *edge fusion*, besides the tree edit operations of deletion, insertion, and relabeling classically used in the literature. This allows us to address some serious limitations of the more traditional tree edit operations when the trees represent RNAs and what is searched for is a common structural core of two RNAs. Although the algorithm complexity has an exponential term, this term depends only on the number of successive fusions that may be applied to a same node, not on the total number of fusions. The algorithm remains therefore efficient in practice and is used for illustrative purposes on ribosomal as well as on other types of RNAs.

**Index Terms**—Tree comparison, edit operation, distance, RNA, secondary structure.

✦

## 1 INTRODUCTION

RNAS are one of the fundamental elements of a cell. Their role in regulation has been recently shown to be far more prominent than initially believed (20 December 2002 issue of *Science*, which designated small RNAs with regulatory function as the scientific breakthrough of the year). It is now known, for instance, that there is massive transcription of noncoding RNAs. Yet current mathematical and computer tools remain mostly inadequate to identify, analyze, and compare RNAs.

An RNA may be seen as a string over the alphabet of nucleotides (also called bases), {A, C, G, T}. Inside a cell, RNAs do not retain a linear form, but instead fold in space. The fold is given by the set of nucleotide bases that pair. The main type of pairing, called canonical, corresponds to bonds of the type $A-U$ and $G-C$. Other rarer types of bonds may be observed, the most frequent among them is $G-U$, also called the wobble pair. Fig. 1 shows the sequence of a folded RNA. Each box represents a consecutive sequence of bonded pairs, corresponding to a helix in 3D space. The secondary structure of an RNA is the set of helices (or the list of paired bases) making up the RNA. Pseudoknots, which may be described as a pair of interleaved helices, are in general excluded from the secondary structure of an RNA. RNA secondary structures can thus be represented as planar graphs. An RNA primary structure is its sequence of nucleotides while its tertiary structure corresponds to the geometric form the RNA adopts in space.

Apart from helices, the other main structural elements in an RNA are:

1. hairpin loops which are sequences of unpaired bases closing a helix;
2. internal loops which are sequences of unpaired bases linking two different helices;
3. bulges which are internal loops with unpaired bases on one side only of a helix;
4. multiloops which are unpaired bases linking at least three helices.

Stems are successions of one or more among helices, internal loops, and/or bulges.

The comparison of RNA secondary structures is one of the main basic computational problems raised by the study of RNAs. It is the problem we address in this paper. The motivations are many. RNA structure comparison has been used in at least one approach to RNA structure prediction that takes as initial data a set of unaligned sequences supposed to have a common structural core [1]. For each sequence, a set of structural predictions are made (for instance, all suboptimal structures predicted by an algorithm like Zucker's MFOLD [15], or all suboptimal sets of compatible helices or stems). The common structure is then found by comparing all the structures obtained from the initial set of sequences, and identifying a substructure common to all, or to some of the sequences. RNA structure comparison is also an essential element in the discovery of RNA structural motifs, or profiles, or of more general models that may then be used to search for other RNAs of the same type in newly sequenced genomes. For instance, general models for tRNAs and introns of group I have been derived by hand [3], [10]. It is an open question whether models at least as accurate as these, or perhaps even more accurate, could have been derived in an automatic way. The identification of smaller structural motifs is an equally important topic that requires comparing structures.

As we saw, the comparison of RNA structures may concern *known* RNA structures (that is, structures that were experimentally determined) or *predicted* structures. The objective in both cases is the same: to find the common parts of such structures.

In [11], Shapiro suggested to mathematically model RNA secondary structures without pseudoknots by means of trees. The trees are rooted and ordered, which means that the order among the children of a node matters. This order corresponds to the 5'-3' orientation of an RNA sequence.

---

- *J. Allali is with the Institut Gaspard-Monge, Université de Marne-la-Vallée, Cité Descartes, Champs-sur-Marne, 77454, Marne-la-Vallée Cedex 2, France. E-mail: allali@univ-mlv.fr.*
- *M.-F. Sagot is with Inria Rhône-Alpes, Université Claude Bernard, Lyon I, 43 Bd du Novembre 1918, 69622 Villeurbanne cedex, France. E-mail: Marie-France.Sagot@inria.fr.*

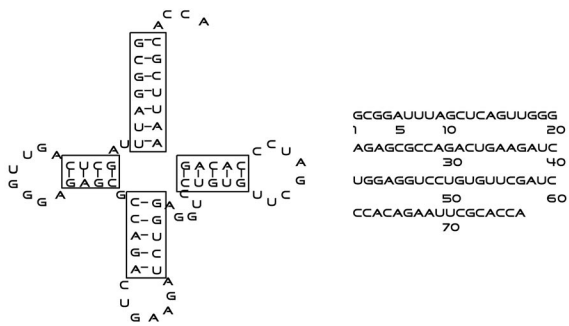

Fig. 1. Primary and secondary structures of a transfer RNA.

Given two trees representing each an RNA, there are two main ways for comparing them. One is based on the computation of the edit distance between the two trees while the other consists in aligning the trees and using the score of the alignment as a measure of the distance between the trees. Contrary to what happens with sequences, the two, alignment and edit distance, are not equivalent. The alignment distance is a restrained form of the edit distance between two trees, where all insertions must be performed before any deletions. The alignment distance for general trees was defined in 1994 by Jiang et al. in [9] and extended to an alignment distance between forests in [6]. More recently, Höchsmann et al. [7] applied the tree alignment distance to the comparison of two RNA secondary structures. Because of the restriction on the way edit operations can be applied in an alignment, we are not concerned in this paper with tree alignment distance and we therefore address exclusively from now on the problem of tree edit distance.

Our way for comparing two RNA secondary structures is then to apply a number of tree edit operations in one or both of the trees representing the RNAs until isomorphic trees are obtained. The currently most popular program using this approach is probably the Vienna package [5], [4]. The tree edit operations considered are derived from the operations classically applied to sequences [13]: substitution, deletion, and insertion. In 1989, Zhang and Shasha [14] gave a dynamic programming algorithm for comparing two trees. Shapiro and Zhang then showed [12] how to use tree editing to compare RNAs. The latter also proposed various tree models that could be used for representing RNA secondary structures. Each suggested tree offers a more or less detailed view of an RNA structure. Figs. 2b, 2c, 2d, and 2e present a few examples of such possible views for the RNA given in Fig. 2a. In Fig. 2, the nodes of the tree in Fig. 2b represent either unpaired bases (leaves) or paired bases (internal nodes). Each node is labelled with, respectively, a base or a pair of bases. A node of the tree in Fig. 2c represents a set of successive unpaired bases or of stacked paired ones. The label of a node is an integer indicating, respectively, the number of unpaired bases or the height of the stack of paired ones. The nodes of the tree in Fig. 2d represent elements of secondary structure: hairpin loop (H), bulge (B), internal loop (I), or multiloop (M). The edges correspond to helices. Finally, the tree in Fig. 2e contains only the information concerning the skeleton of multiloops of an RNA. The last representation, though giving a highly simplified view of an RNA, is important nevertheless as it is generally accepted that it is this skeleton which is usually the most constrained part of an RNA. The last two models may be enriched with information concerning, for instance, the number of (unpaired) bases in a loop (hairpin, internal, multi) or bulge, and the number of paired bases in a helix. The first label the nodes of the tree, the second its edges. Other types of information may be added (such as overall composition of the elements of secondary structure). In fact, one could consider working with various representations simultaneously or in an interlocked, multilevel fashion. This goes beyond the scope of this paper which is concerned with comparing RNA secondary structures using any one among the many tree representations possible. We shall, however, comment further on this multilevel approach later on.

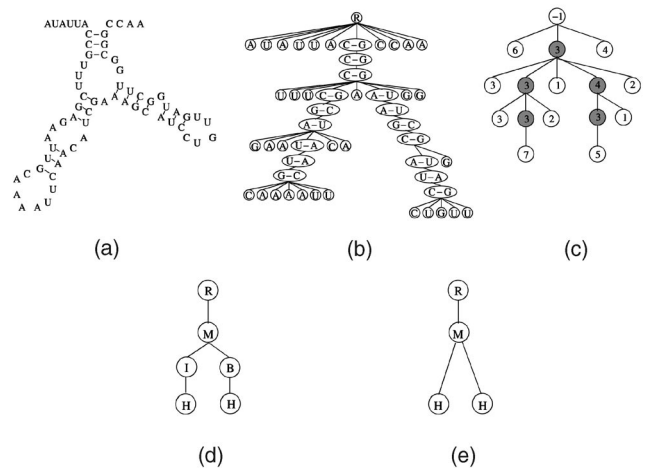

Fig. 2. Example of different tree representations ((b), (c), (d), and (e)) of the same RNA (a).

Concerning the objectives of this paper, they are twofold. The first is to give some indications on why the classical edit operations that have been considered so far in the literature for comparing trees present some limitations when the trees stand for RNA structures. Three cases of such limitations will be illustrated through examples in Section 3. In Section 4, we then introduce two novel operations, so-called *node-fusion* and *edge-fusion*, that enable us to address some of these limitations and then give a dynamic programming algorithm for comparing two RNA structures with these two additional operations. Implementation issues and initial results are presented in Section 4. In Section 5, we give a first application of our algorithm to the comparison of two RNA secondary structures. Finally, in Section 6, we sketch the main ideas behind the multilevel RNA comparison approach mentioned above. Before that, we start by introducing some notation and by recalling in the next section the basics about classical tree edit operations and tree mapping.

This paper is an extended version of a paper presented at the Workshop on Algorithms in BioInformatics (WABI) in 2004, in Bergen, Norway. A few more examples are given to illustrate some of the points made in the WABI paper, complexity and implementation issues are discussed in more depth as are the cost functions and a multilevel approach to comparing RNAs.

## 2 TREE EDITING AND MAPPING

Let $T$ be an ordered rooted tree, that is, a tree where the order among the children of a node matters. We define three kinds of operations on $T$: deletion, insertion, and relabeling (corresponding to a substitution in sequence comparison). The operations are shown in Fig. 3. The deletion (Fig. 3b) of a node $u$ removes $u$ from the tree. The children of $u$ become the children of $u$'s father. An insertion (Fig. 3c) is the symmetric of a deletion. Given a node $u$, we remove a consecutive (in relation to the order among the children) set $u_1, \ldots, u_p$ of its children, create a new node $v$, make $v$ a child of $u$ by attaching it at the place where the set was, and, finally, make the set $u_1, \ldots, u_p$ (in the same order) the children of $v$. The relabeling of a node (Fig. 3d) consists simply in changing its label.

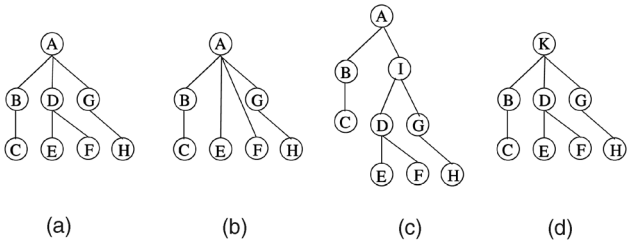
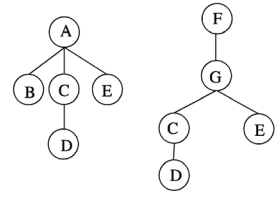

Fig. 3. Edit operations: (a) the original tree $T$, (b) deletion of the node labelled $D$, (c) insertion of the node labeled $I$, and (d) relabeling of a node in $T$ (the label $A$ of the root is changed into $K$).

Given two trees $T$ and $T'$, we define $\mathcal{S} = \{s_1 \ldots s_e\}$ to be a series of edit operations such that, if we apply successively the operations in $\mathcal{S}$ to the tree $T$, we obtain $T'$ (i.e., $T$ and $T'$ become isomorphic). A series of operations like $\mathcal{S}$ realizes the editing of $T$ into $T'$ and is denoted by $T \xrightarrow{\mathcal{S}} T'$.

We define a function $cost$ from the set of possible edit operations (deletion, insertion, relabeling) to the integers (or the reals) such that $cost_s$ is the score of the edit operation $s$. If $\mathcal{S}$ is a series of edit operations, we define by extension that $cost_{\mathcal{S}}$ is $\sum_{s \in \mathcal{S}} cost_s$. We can define the edit distance between two trees as the series of operations that performs the editing of $T$ into $T'$ and such that its cost is minimal: $distance(T, T') = \{\min(cost_{\mathcal{S}}) | T \xrightarrow{\mathcal{S}} T'\}$.

Let an insertion or a deletion cost one and the relabeling of a node cost zero if the label is the same and one otherwise. For the two trees of the figure on the left, the series $relabel(A \to F).delete(B).insert(G)$ realizes the editing of the left tree into the right one and costs 3. Another possibility is the series $delete(B).relabel(A \to G).insert(F)$ which also costs 3. The distance between these two trees is 3.

Given a series of operations $\mathcal{S}$, let us consider the nodes of $T$ that are not deleted (in the initial tree or after some relabeling). Such nodes are associated with nodes of $T'$. The *mapping* $M_{\mathcal{S}}$ relative to $\mathcal{S}$ is the set of couples $(u, u')$ with $u \in T$ and $u' \in T'$ such that $u$ is associated with $u'$ by $\mathcal{S}$.

The operations described above are the "classical tree edit operations" that have been commonly used in the literature for RNA secondary structure comparison. We now present a few results obtained using such classical operations that will allow us to illustrate a few limitations they may present when used for comparing RNA structures.

## 3 LIMITATIONS OF CLASSICAL TREE EDIT OPERATIONS FOR RNA COMPARISON

As suggested in [12], the tree edit operations recalled in the previous section can be used on any type of tree coding of an RNA secondary structure.

Fig. 4 shows two RNAseP extracted from the database [2] (they are found, respectively, in *Streptococcus gordonii* and *Thermotoga maritima*). For the example we discuss now, we code the RNAs using the tree representation indicated in Fig. 2b where a node represents a base pair and a leaf an unpaired base. After applying a few edit operations to the trees, we obtain the result indicated in Fig. 4, with deleted/inserted bases in gray. We have surrounded a few regions that match in the two trees. Bases in the rectangular box at the bottom of the RNA on the left are thus associated with bases in the bottom rightmost rectangular box of the RNA on the right. The same is observed for the bases in the oval boxes for both RNAs. Such matches illustrate one of the main problems with the classical tree edit operations: Bases in one RNA may be mapped to identically labeled bases in the other RNA to minimise the total cost, while such bases should not be associated in terms of the elements of secondary structure to which they belong. In fact, such elements are often distant from one another along the common RNA structure. We call this problem the "scattering effect." It is related to the definition of tree edit operations. In the case of this example and of the representation adopted, the problem might have been avoided if structural information had been used. Indeed, the problem appears also because the structural location of an unpaired base is not taken into account. It is therefore possible to match, for instance, an unpaired base from a hairpin loop with an unpaired base from a multiloop. Using another type of representation, as we shall do, would, however, not be enough to solve all problems as we see next.

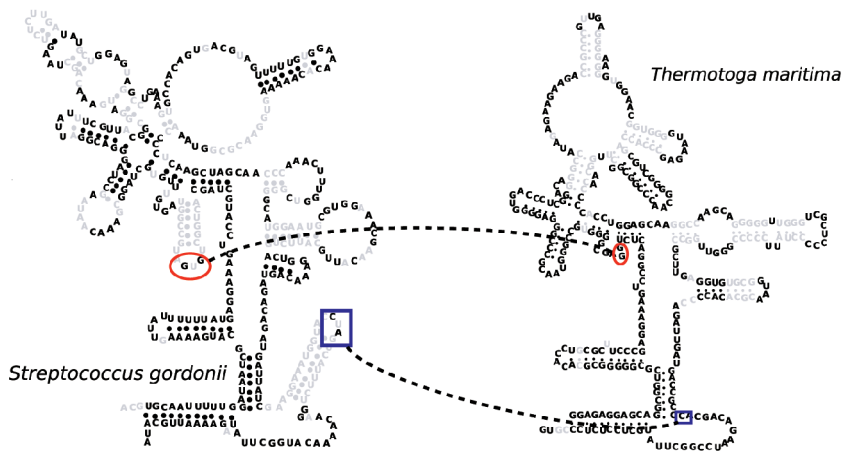

Fig. 4. Illustration of the scattering effect problem. Result of the matching of two RNAseP, of *Streptococcus gordonii* and of *Thermotoga maritima*, using the model given in Fig. 2b.

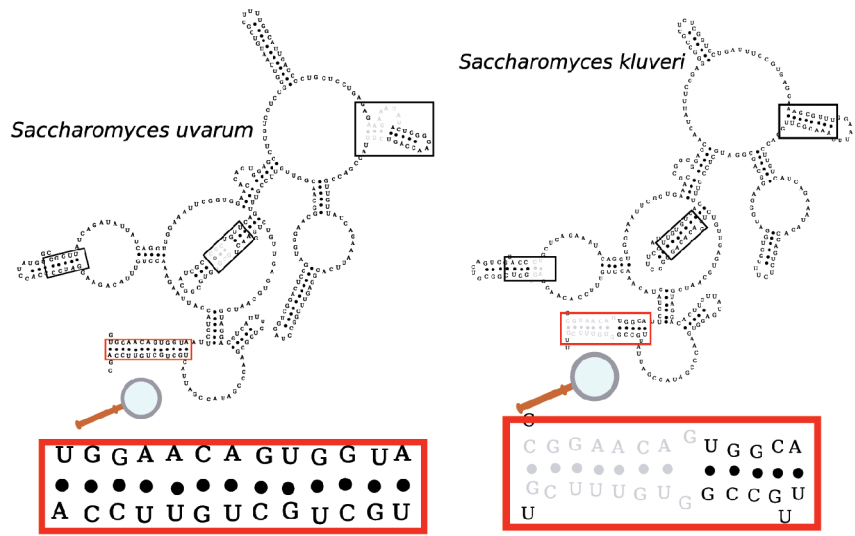

Fig. 5. Illustration of the one-to-one association problem with edges. Result of the matching of the two RNAsePs, of *Saccharomyces uvarum* and of *Saccharomyces kluveri*, using the model given in Fig. 2d.

Indeed, to compare the same two RNAs, we can also use a more abstract tree representation such as the one given in Fig. 2d. In this case, the internal nodes represent a multiloop, internal-loop, or bulge, the leaves code for hairpin loops and edges for helices. The result of the edition of $T$ into $T'$ for some cost function is presented in Fig. 5 (we shall come back later to the cost functions used in the case of such more abstract RNA representations; for the sake of this example, we may assume an arbitrary one is used).

The problem we wish to illustrate in this case is shown by the boxes in the figure. Consider the boxes at the bottom. In the left RNA, we have a helix made up of 13 base pairs. In the right RNA, the helix is formed by seven base pairs followed by an internal loop and another helix of size 5. By definition (see Section 2), the algorithm can only associate one element in the first tree to one element in the second tree. In this case, we would like to associate the helix of the left tree to the two helices of the second tree since it seems clear that the internal loop represents either an inserted element in the second RNA, or the unbonding of one base pair. This, however, is not possible with classical edit operations.

A third type of problem one can meet when using only the three classical edit operations to compare trees standing for RNAs is similar to the previous one, but concerns this time a node instead of edges in the same tree representation. Often, an RNA may present a very small helix between two elements (multiloop, internal-loop, bulge, or hairpin-loop) while such helix is absent in the other RNA. In this case, we would therefore have liked to be able to associate one node in a tree representing an RNA with two or more nodes in the tree for the other RNA. Once again, this is not possible with any of the classical tree edit operations. An illustration of this problem is shown in Fig. 6.

We shall use RNA representations that take the elements of the structure of an RNA into account to avoid some of the scattering effect. Furthermore, in addition to considering information of a structural nature, labels are attached, in general, to both nodes and edges of the tree representing an RNA. Such labels are numerical values (integers or reals). They represent in most cases the size of the corresponding element, but may also further indicate its composition, etc. Such additional information is then incorporated into the cost functions for all three edit operations. It is important to observe that when dealing with trees labeled at both the nodes and edges, any node and the edge that leads to it (or, in an alternative perspective, departs from it) represent a single object from the point of view of computing an edit distance between the trees.

It remains now to deal with the last two problems that are a consequence of the one-to-one associations between nodes and edges enforced by the classical tree edit operations. To that purpose, we introduce two novel tree edit operations, called the *edge fusion* and the *node fusion*.

## 4 INTRODUCING NOVEL TREE EDIT OPERATIONS

### 4.1 Edge Fusion and Node Fusion

In order to address some of the limitations of the classical tree edit operations that were illustrated in the previous section, we need to introduce two novel operations. These are the *edge fusion* and the *node fusion*. They may be applied to any of the tree representations given in Figs. 2c, 2d, and 2e.

An example of edge fusion is shown in Fig. 7a. Let $e_u$ be an edge leading to a node $u$, $c_i$ a child of $u$ and $e_{c_i}$ the edge between $u$ and $c_i$. The edge fusion of $e_u$ and $e_{c_i}$ consists in replacing $e_{c_i}$ and $e_u$ with a *new* single edge $e$. The edge $e$ links the father of $u$ to $c_i$. Its label then becomes a function of the (numerical) labels of $e_u$, $u$ and $e_{c_i}$. For instance, if such labels indicated the size of each element (e.g., for a helix, the number of its stacked pairs, and for a loop, the $\min$, $\max$ or the average of its unpaired bases on each side of the loop), the label of $e$ could be the sum of the sizes of $e_u$, $u$ and $e_{c_i}$. Observe that merging two edges implies deleting all subtrees rooted at the children $c_j$ of $u$ for $j$ different from $i$. The cost of such deletions is added to the cost of the edge fusion.

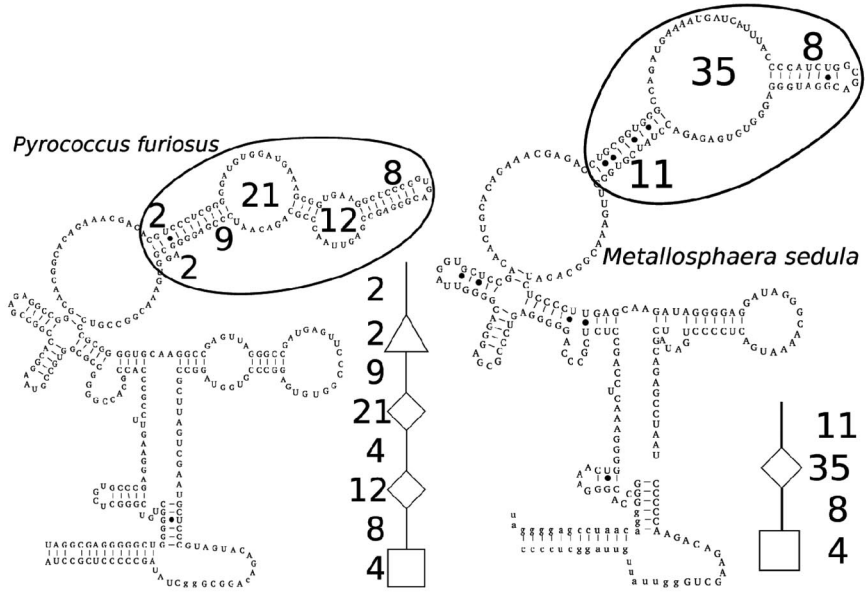

Fig. 6. Illustration of the one-to-one association problem with nodes. The two RNAs used here are RNAsePs from *Pyrococcus furiosus* and *Metallosphaera sedula*. Triangles stand for bulges, diamond stand for internal loops, and squares for hairpin loops.

An example of node fusion is given in Fig. 7b. Let $u$ be a node and $c_i$ one of its children. Performing a node fusion of $u$ and $c_i$ consists in making $u$ the father of all children of $c_i$ and in relabeling $u$ with a value that is a function of the values of the labels of $u$, $c_i$ and of the edge between them.

Observe that a node fusion may be simulated using the classical edit operations by a deletion followed by a relabeling. However, the difference between a node fusion and a deletion/relabeling is in the cost associated with both operations. We shall come back to this point later.

Obviously, like insertions or deletions, edge fusions and node fusions have of course symmetric counterparts, which are the *edge split* and the *node split*.

Given two rooted, ordered, and labeled trees $T$ and $T'$, we define the "edit distance with fusion" between $T$ and $T'$ as $distance_{fusion}(T, T') = \{\min(cost_S) | T \xrightarrow{S} T'\}$ with $cost_s$ the cost associated to each of the seven edit operations now considered (relabeling, insertion, deletion, node fusion and split, edge fusion and split).

**Proposition 1.** *If the following is verified:*

- $cost_{match}(a, b)$ is a distance,
- $cost_{ins}(a) = cost_{del}(a) \geq 0$,
- $cost_{node_{fusion}}(a, b, c) = cost_{node_{split}}(a, b, c) \geq 0$, and
- $cost_{edge_{fusion}}(a, b, c) = cost_{edge_{split}}(a, b, c) \geq 0$,

*then $distance_{fusion}$ is indeed a distance.*

**Proof.** The positiveness of $distance_{fusion}$ is given by the fact that all elementary cost functions are positive. Its symmetry is guaranteed by the symmetry in the costs of the insertion/deletion and (node/edge) fusion/split operations. Finally, it is straighforward to see that $distance_{fusion}$ satisfies triangular inequality. □

Besides the above properties that must be satisfied by the cost functions in order to obtain a distance, others may be introduced for specific purposes. Some will be discussed in Section 5.

We now present an algorithm to compute the tree edit distance between two trees using the classical tree edit operations plus the two operations just introduced.

### 4.2 Algorithm

The method we introduce is a dynamic programming algorithm based on the one proposed by Zhang and Shasha. Their algorithm is divided in two parts: They first compute the edit distance between two trees (this part is denoted by $TDist$) and then the distance between two forests (this part is denoted by $FDist$). Fig. 8 illustrates in pictorial form the part $TDist$ and Fig. 9 the $FDist$ part of the computation.

In order to take our two new operations into account, we need to compute a few more things in the $TDist$ part. Indeed, we must add the possibility for each tree to have a node fusion (inversely, *node split*) between the root and one of its children, or to have an edge fusion (inversely *edge split*) between the root and one of its children. These additional operations are indicated in the right box of Fig. 8.

We present now a formal description of the algorithm. Let $T$ be an ordered rooted tree with $|T|$ nodes. We denote by $t_i$ the $i$th node in a postfix order. For each node $t_i$, $l(i)$ is the index of the leftmost child of the subtree rooted at $t_i$. Let $T(i \ldots j)$ denote the forest composed by the nodes $t_i \ldots t_j$ ($T \equiv T(0 \ldots |T|)$). To simplify notation, from now on, when there is no ambiguity, $i$ will refer to the node $t_i$. In this case, $distance(i_1 \ldots i_2, j_1 \ldots j_2)$ will be equivalent to $distance(T(i_1 \ldots i_2), T'(j_1 \ldots j_2))$.

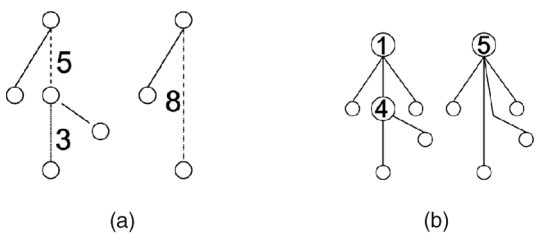

Fig. 7. (a) An example of edge fusion. (b) An example of node fusion.

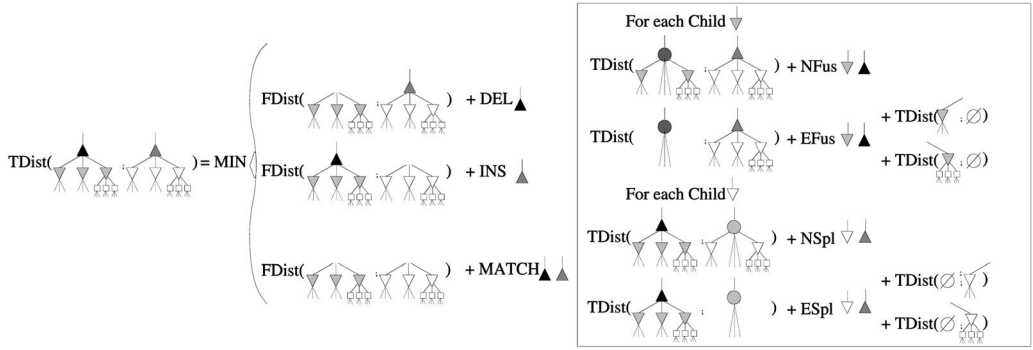

Fig. 8. Zhang and Sasha's dynamic programming algorithm: the tree distance part. The right box corresponds to the additional operations added to take fusion into account.

The algorithm of Zhang and Sasha is fully described by the following recurrence formula:

$$distance(i_1 \ldots i_2, j_1 \ldots j_2) =$$

**if** $((i_1 == l(i_2))$ and $(j_1 == l(j_2)))$

MIN
$$\begin{cases} distance(\ i_1 \ldots i_2-1\ ,\ j_1 \ldots j_2\ ) + cost_{del}(i_2) \\ distance(\ i_1 \ldots i_2\ ,\ j_1 \ldots j_2-1\ ) + cost_{ins}(j_2) \\ distance(\ i_1 \ldots i_2-1\ ,\ j_1 \ldots j_2-1\ ) + cost_{match}(i_2, j_2) \end{cases} \quad (1)$$

**else**

MIN
$$\begin{cases} distance(\ i_1 \ldots i_2-1\ ,\ j_1 \ldots j_2) \\ \quad + cost_{del}(i_2) \\ distance(\ i_1 \ldots i_2)\ ,\ j_1 \ldots j_2-1\ ) \\ \quad + cost_{ins}(j_2) \\ distance(\ i_1 \ldots l(i_2)-1\ ,\ j_1 \ldots l(j_2)-1\ ) \\ \quad + distance(\ l(i_2) \ldots i_2\ ,\ l(j_2) \ldots j_2\ ) \end{cases} \quad (2)$$

Part (1) of the formula corresponds to Fig. 8, while part (2) corresponds to Fig. 9. In practice, the algorithm stores in a matrix the score between each subtree of $T$ and $T'$. The space complexity is therefore $O(|T| * |T'|)$. To reach this complexity, the computation must be done in a certain order (see Section 4.3). The time complexity of the algorithm is

$$O(|T| * min(leaf(T), height(T)) \\ * |T'| * min(leaf(T'), height(T'))),$$

where $leaf(T)$ and $height(T)$ represent, respectively, the number of leaves and the height of a tree $T$.

The formula to compute the edit score allowing for both node and edge fusions follows.

$$distance(\{i_1, \ldots, i_k\}, path, \{j_1, \ldots, j_{k'}\}, path') =$$

**if** $((i_1 \geq l(i_k))$ and $(j_1 \geq l(j_{k'})))$

MIN
$$\begin{cases} distance(\{i_1 \ldots i_{k-1}\}, \emptyset, \{j_1 \ldots j_{k'}\}, path') + cost_{del}(i_k) \\ distance(\{i_1 \ldots i_k\}, path, \{j_1 \ldots j_{k'-1}\}, \emptyset) + cost_{ins}(j_{k'}) \\ distance(\{i_1 \ldots i_{k-1}\}, \emptyset, \{j_1 \ldots j_{k'-1}\}, \emptyset) + cost_{match}(i_k, j_{k'}) \\ \text{for each child } i_c \text{ of } i_k \text{ in } \{i_1, \ldots, i_k\}, \text{ set } i_l = l(i_c) \\ \quad distance(\{i_1 \ldots i_{c-1}, i_{c+1} \ldots i_k\}, path.(u, i_c), \{j_1 \ldots j_{k'}\}, \\ \quad path') \\ \quad\quad + cost_{node\_fusion}(i_c, i_k) (\textbf{obs.} : i_k \text{ data are changed}) \\ \quad distance(\{i_l \ldots i_{c-1}, i_k\}, path.(e, i_c), \{j_1 \ldots j_{k'}\}, path') \\ \quad\quad + cost_{edge\_fusion}(i_c, i_k) + distance(\{i_1 \ldots i_{l-1}\}, \\ \quad\quad \emptyset, \emptyset, \emptyset) \\ \quad\quad + distance(\{i_{c+1} \ldots i_k - 1, \emptyset, \emptyset, \emptyset\}) \\ \quad\quad (\textbf{obs.} : i_k \text{ data are changed}) \\ \text{for each child } j_{c'} \text{ of } j_{k'} \text{ in } \{j_1, \ldots, j_{k'}\}, \text{ set } j_{l'} = l(j_{c'}) \\ \quad distance(\{i_1 \ldots i_k\}, path, \{j_1 \ldots j_{c'-1}, j_{c'+1} \ldots j_{k'}, \\ \quad path'.(u, j_{c'})) \\ \quad\quad + cost_{node\_split}(j_{c'}, j_{k'}) \\ \quad\quad (\textbf{obs.} : j_{k'} \text{ data are changed}) \\ \quad distance(\{i_1 \ldots i_k\}, path, \{j_{l'} \ldots j_{c'}, j_{k'}, path'.(e, j_{c'})\}) \\ \quad\quad + cost_{edge\_split}(j_{c'}, j_{k'}) \\ \quad\quad + distance(\emptyset, \emptyset, \{j_1 \ldots j_{l'-1}\}, \emptyset) \\ \quad\quad + distance(\emptyset, \emptyset, j_{c'+1} \ldots j_{k'-1}, \emptyset) \\ \quad\quad (\textbf{obs.} : j_{k'} \text{ data are changed}) \end{cases}$$
$$(3)$$

**else** set $i_{l'} = l(i_k)$ and $j_{l'} = l(j_{k'})$

MIN
$$\begin{cases} distance(\{i_1 \ldots i_{k-1}\}, \emptyset, \{j_1 \ldots j_{k'}\}, path') + del(i_k) \\ distance(\{i_1 \ldots i_k\}, path, \{j_1 \ldots j_{k'-1}\}, \emptyset) + ins(j_{k'}) \quad (4) \\ distance(\{i_1 \ldots i_{l-1}\}, \emptyset, \{j_1 \ldots j_{l'-1}\}, \emptyset) \\ \quad + distance(\{i_l \ldots i_k\}, path, \{j_{l'} \ldots j_{k'}\}, path') \end{cases}$$

Given two nodes $u$ and $v$ such that $v$ is a child of $u$, $node\_fusion(u, v)$ is the fusion of node $v$ with $u$, and $edge\_fusion(u, v)$ is the edge fusion between the edges leading to, respectively, nodes $u$ and $v$. The symmetric operations are denoted by, respectively, $node\_split(u, v)$ and $edge\_split(u, v)$.

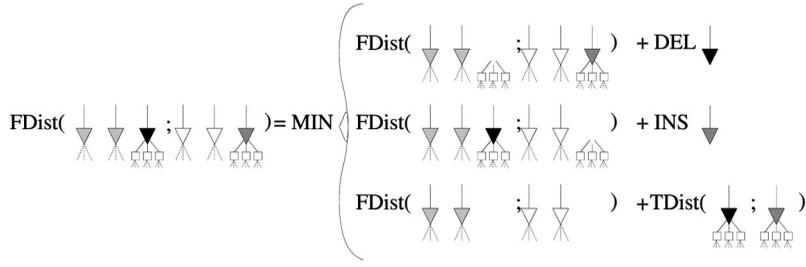

Fig. 9. Zhang and Sasha's dynamic programming algorithm: the forest distance part.

The distance computation takes two new parameters $path$ and $path'$. These are sets of pairs ($e$ or $u, v$) which indicate, for node $i_k$ (respectively, $j_k$), the series of fusions that were done. Thus, a pair ($e, v$) indicates that an edge fusion has been perfomed between $i_k$ and $v$, while for ($u, v$) a node $v$ has been merged with node $i_k$.

The notation $path.(e, v)$ indicates that the operation ($e, v$) has been performed in relation to node $i_k$ and the information is thus concatenated to the set $path$ of pairs currently linked with $i_k$.

### 4.3 Implementation and Complexity

The previous section gave the recurrence formulæ for calculating the edit distance between two trees allowing for node and edge fusion and split. We now discuss the complexity of the algorithm. This requires paying attention to some high-level implementation details that, in the case of the tree edit distance problem, may have an important influence on the theoretical complexity of the algorithm. Such details were first observed by Zhang and Shasha. They concern the order in which to perform the operations indicated in (2) and (1) to obtain an algorithm that is time and space efficient.

Let us consider the last line of (2). We may observe that the computation of the distance between two forests refers to the computation of the distance between two trees $T(l(i_2)\ldots i_2)$ and $T'(l(j_2)\ldots j_2)$. We must therefore memorise the distance between any two subtrees of $T$ and $T'$. Furthermore, we have to carry out the computation from the leaves to the root because when we compute the distance between two subtrees $U$ and $U'$, the distance between any subtrees of $U$ and $U'$ must already have been measured. This explains the space complexity which is in $O(|T| * |T'|)$ and corresponds to the size of the table used for storing such distances in memory.

If we look at (1) now, we see that it is not necessary to calculate separately the distance between the subtrees rooted at $i'$ and $j'$ if $i'$ is on the path from $l(i)$ to $i$ and $j'$ is on the path from $l(j)$ to $j$, for $i$ and $j$ nodes of, respectively, $T$ and $T'$.

We define a set $LR(T)$ of the left roots of $T$ as follows:

$LR(T) = \{k | 1 \leq k \leq |T| \text{ and } \nexists k' > k \text{ such that } l(k') = l(k)\}$

The algorithm for computing the edit distance between $t$ and $T'$ consists then in computing the distance between each subtree rooted at a node in $LR(T)$ and each subtree rooted at a node in $LR(T')$. Such subtrees are considered from the leaves to the root of $T$ and $T'$, that is, in the order of their indexes.

Zhang and Shasha proved that this algorithm has a time complexity in $O(|T| * min(leaf(T), height(T)) * |T'| * min(leaf(T'), height(T')))$, $leaf(T)$ designating the number of leaves of $T$ and $height(T)$ its height. In the worst case (fan tree), the complexity is in $O(|T|^2 * |T'|^2)$.

Taking fusion and split operations into account does not change the above reasoning. However, we must now store in memory the distance between all subtrees $T(l(i_2)\ldots i_2)$ and $T'(l(j_2)\ldots j_2)$, and all the possible values of $path$ and $path'$.

We must therefore determine the number of values that $path$ can take. This amounts to determine the total number of successive fusions that could be applied to a given node. We recall that $path$ is a list of pairs ($e$ or $u, v$). Let $path = \{(e \text{ or } u, v_1), (e \text{ or } u, v_2), \ldots, (e \text{ or } u, v_\ell)\}$ be the list for node $i$ of $T$. The first fusion can be performed only with a child $v_1$ of $i$. If $d$ is the maximum degree of $T$, there are $d$ possible choices for $v_1$. The second fusion can be done with one of the children of $i$ or with one of its grandchildren. Let $v_2$ be the node chosen. There are $d + d^2$ possible choices for $v_2$. Following the same reasoning, there are $\sum_{k=1}^{k=\ell} d^k$ possible choices for the $\ell$th node $v_\ell$ to be fusioned with $i$.

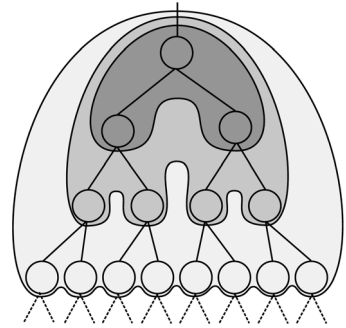

Furthermore, we must take into account the fact that a fusion can concern a node or an edge. The total number of values possible for the variable $path$ is therefore:

$$2^\ell * \prod_{k=1}^{k=\ell} \sum_{j=1}^{j=k} d^j = 2^l \prod_{k=1}^{k=\ell} \frac{d^{k+1}-1}{d-1},$$

that is:

$$2^\ell * \left(\frac{1}{d-1}\right)^\ell \prod_{k=1}^{k=\ell}(d^{k+1}-1) < 2^l * \left(\frac{1}{d-1}\right)^l * d^{\frac{(\ell+1)(\ell+2)}{2}}.$$

A node $i$ may then be involved in $O((2d)^l)$ possible successive (node/edge) fusions.

As indicated, we must store in memory the distance between each subtree $T(l(i_2)\ldots i_2)$ and $T'(l(j_2)\ldots j_2)$ for all possible values of $path$ and $path'$. The space complexity of our algorithm is thus in $O((2d)^\ell * (2d')^\ell * |T| * |T'|)$, with $d$ and $d'$ the maximum degrees of, respectively, $T$ and $T'$.

The computation of the time complexity of our algorithm is done in a similar way as for the algorithm of Zhang and Shasha. For each node of $T$ and $T'$, one must compute the number of subtree distance computations the node will be involved in by considering all subtrees rooted in, respectively, a node of $LR(T)$ and a node of $LR(T')$. In our case, one must also take into account for each node the possibility of applying a fusion. This leads to a time complexity in

$O((2d)^\ell * |T| * min(leaf(T), height(T)) * (2d')^\ell * |T'| * min(leaf(T'), height(T')))$.

This complexity suggests that the fusion operations may be used only for reasonable trees (typically, less than 100 nodes) and small values of $l$ (typically, less than 4). It is however important to observe that the overall number of fusions one may perform can be much greater than $l$ without affecting the worst-case complexity of the algorithm. Indeed, any number of fusions can be made while still retaining the bound of

$O((2d)^l * |T| * min(leaf(T), height(T)) * |T'| * min(leaf(T'), height(T')))$

so long as one does not realize more than $l$ consecutive fusions for each node.

In general, also, most interesting tree representations of an RNA are of small enough size as will be shown next, together with some initial results obtained in practice.

## 5  APPLICATION TO RNA SECONDARY STRUCTURES COMPARISON

The algorithm presented in the previous section has been coded using C++. An online version is available at http://www-igm.univ-mlv.fr/~allali/migal/.

We recall that RNAs are relatively small molecules with sizes limited to a few kilobases. For instance, the small ribosomal subunit of *Sulfolobus acidocaldarius* (D14876) is made up of 1,147 bases. Using the representation shown in Fig. 2b, the tree obtained contains 440 internal nodes and 567 leaves, that is 1,007 nodes overall. Using the representation in Fig. 2d, the tree is composed of 78 nodes. Finally, the tree obtained using the representation given in Fig. 2e contains only 48 nodes. We therefore see that even for large RNAs, any of the known abstract tree-representations (that is, representations which take the elements of the secondary structure of an RNA into account) that we can use leads to a tree of manageable size for our algorithm. In fact, for small values of $l$ (2 or 3), the tree comparison takes reasonable time (a few minutes) and memory (less than 1Gb).

As we already mentioned, a fusion (respctively, split) can be viewed as an alternative to a deletion (respectively, insertion) followed by a relabeling. Therefore, the cost function for a fusion must be chosen carefully.

To simplify, we reason on the cost of a node fusion without considering the label of the edges leading to the nodes that are fusioned with a father. The formal definition of the cost functions takes the edges also into account.

Let us assume that the cost function returns a real value between zero and one. If we want to compute the cost of a fusion between two nodes $u$ and $v$, the aim is to give to such fusion a cost slightly greater than the cost of deleting $v$ and relabeling $u$; that is, we wish to have $cost_{node\_fusion}(u,v) = min(cost_{del}(v) + t, 1)$. The parameter $t$ is a *tuning parameter* for the fusion.

Suppose that the new node $w$ resulting from the fusion of $u$ and $v$ matches with another node $z$. The cost of this match is $cost_{match}(w,z)$. If we do not allow for node fusions, the algorithm will first match $u$ with $z$, then will delete $v$. If we compare the two possibilities, on one hand we have a total cost of $cost_{node\_fusion}(u,v) + cost_{match}(w,z)$ for the fusion, that is, $cost_{del}(v) + t + cost_{match}(w,z)$, on the other hand, a cost of $cost_{del}(v) + cost_{match}(u,z)$. Thus, $t$ represents the gain that must be obtained by $cost_{match}(w,z)$ with regard to $cost_{match}(u,z)$, that is, by a match without fusion. This is illustrated in Fig. 10.

In this example, the cost associated with the path on the top is $cost_{match}(5,9) + cost_{del}(3)$. The path at the bottom has a cost of $cost_{node\_fusion}(5,3) = cost_{del}(3) + t$ for the node fusion to which is added a relabeling cost of $cost_{match}(8,9)$, leading to a total of $cost_{match}(8,9) + cost_{del}(3) + t$. A node fusion will therefore be chosen if $cost_{match}(8,9) + t > cost_{match}(5,9)$, therefore if the score of a match with fusion is better by at least $t$ than a match without fusion.

We apply the same reasoning to the cost of an edge fusion. The cost function for a node and an edge fusion between a node $u$ and a node $v$, with $e_u$ denoting the edge leading to $u$ and $e_v$ the edge leading to $v$ is defined as follows:

$$cost_{node\_fusion}(u,v) = cost_{del}(v) + cost_{del}(e_v) + t$$
$$cost_{edge\_fusion}(u,v) = cost_{del}(u) + cost_{del}(e_u) + t$$
$$+ \sum_{c\, sibling\, of\, v} \text{cost deleting subtree rooted at } c.$$

The tuning parameter $t$ is thus an important parameter that allows us to control fusions. Always considering a cost function that produces real values between 0 and 1, if $t$ is equal to $0.1$, a fusion will be performed only if it improves the score by $0.1$. In practice, we use values of $t$ between 0 and 0.2.

For practical considerations, we also set a further condition on the cost and relabeling functions related to a node or edge resulting from a fusion which is as follows:

$$cost_{del}(a) + cost_{del}(b) \geq cost_{del}(c)$$

with $c$ the label of the node/edge resulting from the fusion of the nodes/edges labeled $a$ and $b$. Indeed, if this condition is not fulfilled, the algorithm may systematically fusion the nodes or edges to reduce the overall cost.

An important consequence of the conditions seen above is that a node fusion cannot be followed by an edge fusion. Below, the node fusion followed by an edge fusion costs:

$$(cost_{del}(b) + cost_{del}(B) + t) + (cost_{del}(AB) + cost_{del}(a) + t).$$

The alternative is to destroy node $B$ (together with edge $b$) and then to operate an edge fusion, the whole costing: $(cost_{del}(b) + cost_{del}(B)) + (cost_{del}(A) + cost_{del}(a) + t)$. The difference be-tween these two costs is $t + cost_{del}(AB) - cost_{del}(A)$, which is always positive.

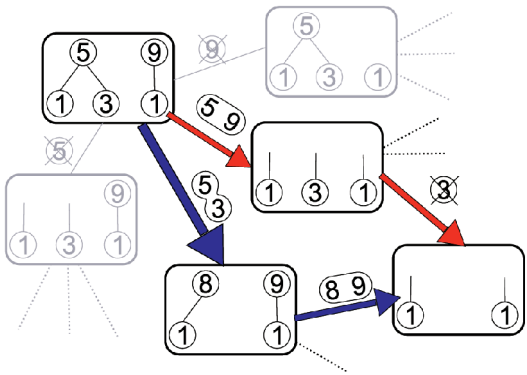

Fig. 10. Illustration of the gain that must be obtained using a fusion instead of a deletion/relabeling.

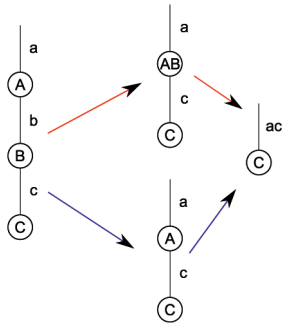

This observation allows to significantly improve the performance in practice of the algorithm.

We have applied the new algorithm on the two RNAs shown in Fig. 5 (these are eukaryotic nuclear P RNAs from *Saccharomyces uvarum* and *Saccharomyces kluveri*) and coded using the same type of representation as in Fig. 2d. We have limited the number of consecutive fusions to one ($l = 1$). The computation of the edit distance between the two trees taking node and edge fusions into account besides deletions, insertions, and relabeling has required less than a second. The total cost allowing for fusions is 6.18 with $t = 0.05$ against 7.42 without fusions. As indicated in Fig. 11, the last two problems discussed in Section 3 disappear thanks to some edge fusions (represented by the boxes).

An example of node fusions required when comparing two "real" RNAs is given in Fig. 12. The RNAs are coded using the same type of representation as in Fig. 2d. The figure shows part of the mapping obtained between the small subunits of two ribosomal RNAs retrieved from [8] (from *Bacillaria paxillifer* and *Calicophoron calicophorum*). The node fusion has been circled.

## 6 MULTILEVEL RNA STRUCTURE COMPARISON: SKETCH OF THE MAIN IDEA

We briefly discuss now an approach which addresses in part the "scattering effect" problem (see Section 2). This approach is being currently validated and will be more fully described in another paper. We therefore present here the main idea only.

To start with, it is important to understand the nature of this "scattering effect." Let us consider first a trivial case: the cost functions are unitary (insertion, deletion, and relabeling each cost 1) and we compute the edit distance between two trees composed of a single node each. The obtained mapping will associate the single node in the first tree with the single one in the second tree, independently from the labels of the nodes. This example can be extended to the comparison of two trees whose node labels are all different. In this case, the obtained mapping corresponds to the maximum homeomorphic subtree common to both trees.

If the two RNA secondary structures compared using a tree representation which models both the base pairs and the nonpaired bases are globally similar but present some local dissimilarity, then an edit operation will almost always associate the nodes of the locally divergent regions that are located at the same positions relatively to the global common structure. This is a normal, expected behavior in the context of an edition. However, it seems clear also when we look at Fig. 4 that the bases of a terminal loop should not be mapped to those of a multiple loop.

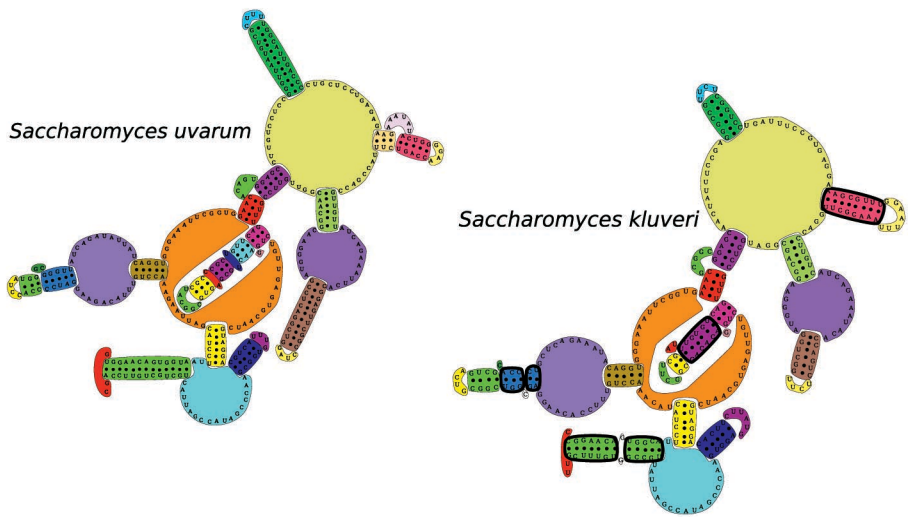

Fig. 11. Result of the editing between the two RNAs shown in Fig. 4 allowing for node and edge fusions.

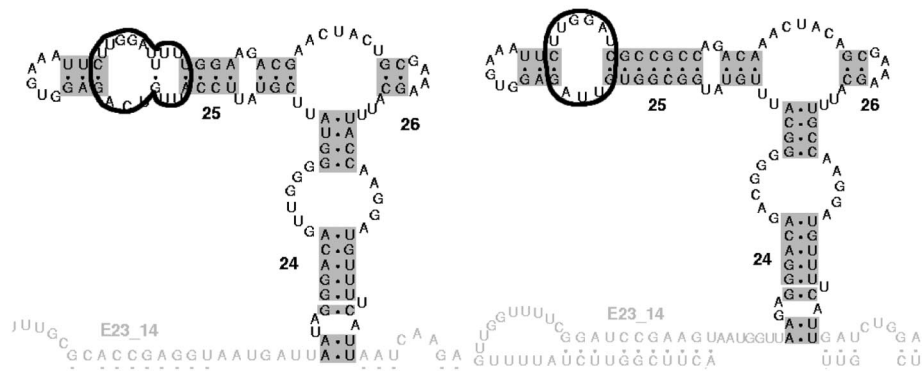

Fig. 12. Part of a mapping between two rRNA small subunits. The node fusion is circled.

To reduce this problem, one possible solution consists of adding to the nodes corresponding to a base an information concerning the element of secondary structure to which the base belongs. The cost functions are then adapted to take this type of information into account. This solution, although producing interesting results, is not entirely satisfying. Indeed, the algorithm will tend to systematically put into correspondence nodes (and, thus, bases) belonging to structural elements of the same type, which is also not necessarily a good choice as these elements may not be related in the overall structure. It seems therefore preferable to have a structural approach first, mapping initially the elements of secondary structure to each other and taking care of the nucleotides in a second step only.

The approach we have elaborated may be briefly described as follows: Given two RNA secondary structures, the first step consists in coding the RNAs by trees of type ($c$) in Fig. 2 (nodes represent bulges or multiple, internal or terminal loops while edges code for helices). We then compute the edit distance between these two trees using the two novel fusion operations described in this paper. This also produces a mapping between the two trees. Each node and edge of the trees, that is, each element of secondary structure, is then colored according to this mapping. Two elements are thus of a same color if they have been mapped in the first step. We now have at our disposal an information concerning the structural similarity of the two RNAs. We can then code the RNAs using a tree of type ($b$). To these trees, we add to each node the colour of the structural element to which it belongs. We need now only to restrict the match operation to nodes of the same color. Two nodes can therefore match only if they belong to secondary elements that have been identified in the first step as being similar.

To illustrate the use of this algorithm, we have applied it to the two RNAs of Fig. 4. Fig. 13 presents the trees of type (Fig. 2c) coding for these structures, and the mapping produced by the computation of the edit distance with fusion. In particular, the noncolored fine dashed nodes and edges correspond, respectively, to deleted nodes/edges. One can see that in the left RNA, the two hairpin loops involved in the scattering effect problem in Fig. 4 (indicated by the arrows) have been destroyed and will not be mapped to one another anymore when the edit operations are applied to the trees of the type in Fig. 2b.

This approach allows to obtain interesting results. Furthermore, it considerably reduces the complexity of the algorithm for comparing two RNA structures coded with trees of the type in Fig. 2b. However, it is important to observe that the scattering effect problem is not specific of the tree representations of the type in Fig. 2b. Indeed, the same problem may be observed, to a lesser degree, with trees of the type in Fig. 2c. This is the reason why we generalize the process by adopting a modelling of RNA secondary structures at different levels of abstraction. This model, and the accompanying algorithm for comparing RNA structures, is in progress.

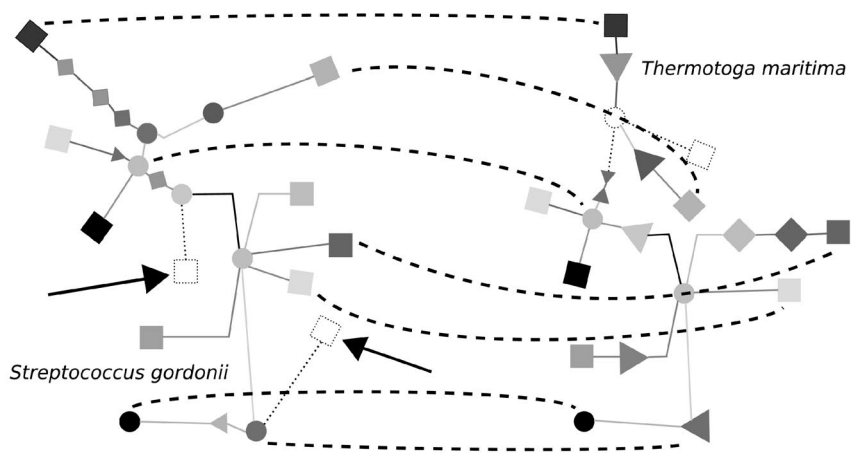

Fig. 13. Result of the comparison of the two RNAs of Fig. 4 using trees in Fig. 2c. The thick dash lines indicate some of the associations resulting from the computation of the edit distance between these two trees. Triangular nodes stand for bulges, diamonds for internal loops, squares for hairpin loops, and circles for multiloops. Noncolored fine dashed nodes and lines correspond, respectively, to deleted nodes/edges.

## 7 FURTHER WORK AND CONCLUSION

We have proposed an algorithm that addresses two main limitations of the classical tree edit operations for comparing RNA secondary structures. Its complexity is high in theory if many fusions are applied in succession to any given (the same) node, but the total number of fusions that may be performed is not limited. In practice, the algorithm is fast enough for most situations one can meet in practice.

To provide a more complete solution to the problem of the scattering effect, we also proposed a new multilevel approach for comparing two RNA secondary structures whose main idea was sketched in this paper. Further details and evaluation of such novel comparison scheme will be the subject of another paper.

## REFERENCES

[1] D. Bouthinon and H. Soldano, "A New Method to Predict the Consensus Secondary Structure of a Set of Unaligned RNA Sequences," *Bioinformatics,* vol. 15, no. 10, pp. 785-798, 1999.
[2] J.W. Brown, "The Ribonuclease P Database," *Nucleic Acids Research,* vol. 24, no. 1, p. 314, 1999.
[3] N. el Mabrouk and F. Lisacek, "and Very Fast Identification of RNA Motifs in Genomic DNA. Application to tRNA Search in the Yeast Genome," *J. Molecular Biology,* vol. 264, no. 1, pp. 46-55, 1996.
[4] I. Hofacker, "The Vienna RNA Secondary Structure Server," 2003.
[5] I. Hofacker, W. Fontana, P.F. Stadler, L. Sebastian Bonhoeffer, M. Tacker, and P. Schuster, "Fast Folding and Comparison of RNA Secondary Structures," *Monatshefte für Chemie,* vol. 125, pp. 167-188, 1994.
[6] M. Höchsmann, T. Töller, R. Giegerich, and S. Kurtz, "Local Similarity in RNA Secondary Structures," *Proc. IEEE Computer Soc. Conf. Bioinformatics,* p. 159, 2003.
[7] M. Höchsmann, B. Voss, and R. Giegerich, "Pure Multiple RNA Secondary Structure Alignments: A Progressive Profile Approach," *IEEE/ACM Trans. Computational Biology and Bioinformatics,* vol. 1, no. 1, pp. 53-62, 2004.
[8] T. Winkelmans, J. Wuyts, Y. Van de Peer, and R. De Wachter, "The European Database on Small Subunit Ribosomal RNA," *Nucleic Acids Research,* vol. 30, no. 1, pp. 183-185, 2002.
[9] T. Jiang, L. Wang, and K. Zhang, "Alignment of Trees—An Alternative to Tree Edit," *Proc. Fifth Ann. Symp. Combinatorial Pattern Matching,* pp. 75-86, 1994.
[10] F. Lisacek, Y. Diaz, and F. Michel, "Automatic Identification of Group I Intron Cores in Genomic DNA Sequences," *J. Molecular Biology,* vol. 235, no. 4, pp. 1206-1217, 1994.
[11] B. Shapiro, "An Algorithm for Multiple RNA Secondary Structures," *Computer Applications in the Biosciences,* vol. 4, no. 3, pp. 387-393, 1988.
[12] B.A. Shapiro and K. Zhang, "Comparing Multiple RNA Secondary Structures Using Tree Comparisons," *Computer Applications in the Biosciences,* vol. 6, no. 4, pp. 309-318, 1990.
[13] K.-C. Tai, "The Tree-to-Tree Correction Problem," *J. ACM,* vol. 26, no. 3, pp. 422-433, 1979.
[14] K. Zhang and D. Shasha, "Simple Fast Algorithms for the Editing Distance between Trees and Related Problems," *SIAM J. Computing,* vol. 18, no. 6, pp. 1245-1262, 1989.
[15] M. Zuker, "Mfold Web Server for Nucleic Acid Folding and Hybridization Prediction," *Nucleic Acids Research,* vol. 31, no. 13, pp. 3406-3415, 2003.

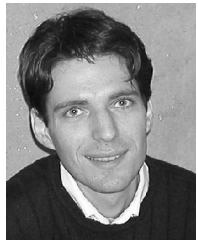

**Julien Allali** studied at the University of Marne la Vallée (France), where he received the MSc degree in computer science and computational genomics. In 2001, he began his PhD in computational genomics at the Gaspard Monge Institute of the University of Marne la Vallée. His thesis focused on the study of RNA secondary structures and, in particular, their comparison using a tree distance. In 2004, he received the PhD degree.

**Marie-France Sagot** received the BSc degree in computer science from the University of São Paulo, Brazil, in 1991, the PhD degree in theoretical computer science and applications from the University of Marne-la-Vallée, France, in 1996, and the Habilitation from the same university in 2000. From 1997 to 2001, she worked as a research associate at the Pasteur Institute in Paris, France. In 2001, she moved to Lyon, France, as a research associate at the INRIA, the French National Institute for Research in Computer Science and Control. Since 2003, she has been the Director of Research at the INRIA. Her research interests are in computational biology, algorithmics, and combinatorics.